\documentclass[reprint, pra, showpacs]{revtex4-1}
\usepackage{hyperref}
\usepackage{graphicx}
\usepackage{amsmath}

\begin{document}

% From braket.sty

\def\bra#1{\mathinner{\langle{#1}|}}
\def\ket#1{\mathinner{|{#1}\rangle}}
\def\braket#1{\mathinner{\langle{#1}\rangle}}
\def\Bra#1{\left<#1\right|}
\def\Ket#1{\left|#1\right>}
{\catcode`\|=\active 
  \gdef\Braket#1{\left<\mathcode`\|"8000\let|\BraVert {#1}\right>}}
\def\BraVert{\egroup\,\mid@vertical\,\bgroup}
% The \mid@vertical is \vrule with ordinary TeX but \middle| in eTeX.
% We always avoid a \mathchoice in making the inner vertical lines.  
% Note that \right>, prints the same as \right\rangle but is faster.  
%
\def\ketbra#1#2{\ket{#1}\bra{#2}}
\def\Ketbra#1#2{\left|{#1}\vphantom{#2}\right>\left<{#2}\vphantom{#1}\right|}

% \Set{...|...} Only the first | is treated specially.
{\catcode`\|=\active
  \gdef\set#1{\mathinner{\lbrace\,{\mathcode`\|"8000\let|\midvert #1}\,\rbrace}}
  \gdef\Set#1{\left\{\:{\mathcode`\|"8000\let|\SetVert #1}\:\right\}}}
\def\midvert{\egroup\mid\bgroup}
\def\SetVert{\egroup\;\mid@vertical\;\bgroup}

% % If the user is using e-TeX with its \middle primitive, use that for
% % verticals instead of \vrule.
% %
% \begingroup
%  \edef\@tempa{\meaning\middle}
%  \edef\@tempb{\string\middle}
% \expandafter \endgroup \ifx\@tempa\@tempb
%  \def\mid@vertical{\middle|}
% \else
%  \let\mid@vertical\vrule
% \fi
% End from braket.sty

% Units
\newcommand{\um}{\ensuremath{\mu \mathrm{m}}}
\newcommand{\uK}{\ensuremath{\mu \mathrm{K}}}
\newcommand{\us}{\ensuremath{\mu \mathrm{s}}}
\newcommand{\nm}{\ensuremath{\mathrm{nm}}}
\newcommand{\mm}{\ensuremath{\mathrm{mm}}}
\newcommand{\ms}{\ensuremath{\mathrm{ms}}}
\newcommand{\W}{\ensuremath{\mathrm{W}}}
\newcommand{\mW}{\ensuremath{\mathrm{mW}}}
\newcommand{\ppm}{\ensuremath{\mathrm{ppm}}}

\newcommand{\HH}{\ensuremath{\mathrm{H}}}
\newcommand{\HHI}{\ensuremath{\HH_\mathrm{I}}}
\newcommand{\cavityfreq}{\omega_\mathrm{C}}
\newcommand{\laserfreq}{\omega_\mathrm{L}}
\newcommand{\Fprimefreq}{\omega_\mathrm{F'}}
\newcommand{\Dperp}{ \ensuremath{D_i^\perp(2, F')}}
\newcommand{\Dpi}{ \ensuremath{D_i^0(2, F')}}
\newcommand{\MHz}{\ensuremath{\mathrm{MHz}}}
\newcommand{\kHz}{\ensuremath{\mathrm{kHz}}}
\providecommand{\abs}[1]{\lvert#1\rvert}
\newcommand{\gbar}{\ensuremath{\bar{g}}}
\newcommand{\gbarFprime}{\ensuremath{\gbar_{F'}}}
\newcommand{\Rb}{\ensuremath{^{87}\mathrm{Rb}}}
\newcommand{\ClebschPi}{\ensuremath{\braket{F, m_F | \mu_0 | F',
m_F}}}

\newcommand{\nempty}{\ensuremath{\bar{n}_\text{empty}}}

\title{Collective cavity QED with multiple atomic levels}
\author{Kyle J. Arnold$^\dagger$}
\author{Markus P. Baden$^\dagger$}
\author{Murray D. Barrett}
\email{phybmd@nus.edu.sg}
\affiliation{$^\dagger$These authors contributed equally to this
    work.\\
    Centre for Quantum Technologies and Department of
  Physics, National University of Singapore, 3 Science Drive 2, 117543 Singapore}

\begin{abstract}
  We study the transmission spectra of ultracold Rubidium atoms
  coupled to a high-finesse optical cavity. Under weak probing with
  $\pi$-polarized light, the linear response of the system is that of
  a collective spin with multiple levels coupled to a single mode of
  the cavity. By varying the atom number, we change the collective
  coupling of the system. We observe the change in transmission
  spectra when going from a regime where the collective coupling is
  much smaller than the separation of the atomic levels to a regime
  where both are of comparable size. The observations are in good
  agreement with a reduced model we developed for our system.
\end{abstract}


\maketitle

\section{Introduction}
\label{sec:introduction}

The coherent interaction between light and matter in the context of
cavity quantum electrodynamics (cavity QED) has been under intensive
study in recent years, especially in the context of quantum
information processing \cite{vanenk:04}. The essence of this
interaction is captured by considering a two-level atom coupled to a
single mode of the electromagnetic field inside a cavity
\cite{jaynes:63,haroche:06}. However, for many physical systems, two
important extensions to this model have to be taken into
account. First, real scatterers often have a more complex level
structure \cite{birnbaum:06} and second, the coupling for $N$
scatterers coupled to the same cavity mode is collectively enhanced
\cite{tavis:68}. Both extensions provide useful functionality in the
context of quantum information processing. The collective enhancement
has recently been used to store multiple microwave pulses in the
collective modes of an electron spin ensemble \cite{wu:10}, whereas
different levels could be used to collectively encode several qubits
\cite{pederson:09}, resulting in quantum repeaters able to perform
simple error correction~\cite{jiang:07}.

Studying a system of $N$ alkali-metal atoms coupled to a cavity allows one
to study both effects in detail. When probing the system with
$\pi$-polarized light for example, the transmission spectrum deviates
strongly from the simple two-level picture. For a two-level atom, the
spectrum shows a single splitting due to an avoided crossing between
the energies of the state with one excitation in the atom and the
state with one excitation in the cavity \cite{boca:04}. Alkali-metal atoms
have multiple hyperfine ground and excited states. In the spectrum,
there are avoided crossings associated with the transitions between
the different ground and excited states \cite{birnbaum:06} and the net
spectrum depends on the relative strength of the transitions and their
separation. For $N$ atoms the shape of the spectrum depends on the
atom number $N$ as well, because the size of the splittings that
correspond to the different avoided crossings is collectively enhanced
by a factor $\sqrt{N}$, as has been observed for thermal atoms
\cite{raizen:89, bernardot:92, thompson:98, tuchman:06} and for
Bose-Einstein condensates \cite{brennecke:07,colombe:07}.

\begin{figure}[tb]
  \centering
  \includegraphics[width=8.6cm]{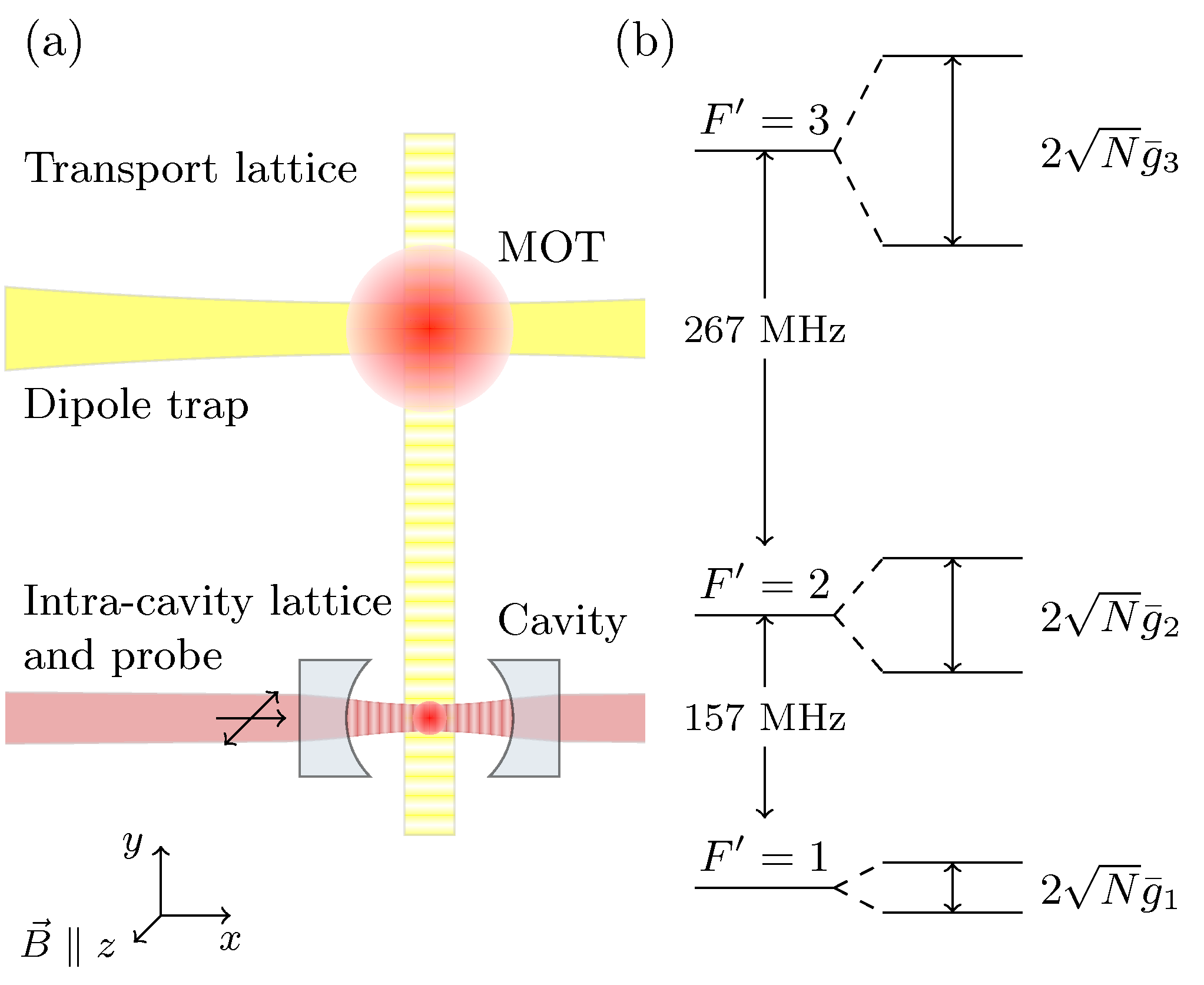}
  \caption{Setup of the experiment. (a) We trap an unpolarized gas of
    ultracold $\Rb$ atoms in a two-dimensional lattice inside the mode
    of a high-finesse optical cavity and probe the system weakly with
    $\pi$-polarized light. (b) Under weak driving the system behaves
    as a collective spin with three transitions coupled to a single
    mode of the cavity. Of interest are the transitions to the excited
    states $\ket{F'=1,2,3}$ of the $D2$ line. The corresponding
    splittings are separated by the hyperfine splitting of the excited
    states. Their size is given by the collective coupling, which in
    turn depends on the atom number $N$ and an effective single atom
    coupling constant $\gbar_{F'}$. By varying $N$, we change the
    collective coupling and are able to explore different regimes of
    cavity QED with multiple atomic levels.}
  \label{fig:sketch}
\end{figure}

Here, we experimentally investigate the transmission spectra of $N$
$\Rb$ atoms coupled to a high-finesse optical cavity when the system
is weakly probed with $\pi$-polarized light. By changing the atom
number, we are able to go from a regime where the size of the
splittings is much smaller than the separation of the hyperfine
excited states to a regime where both are of comparable
size. Depending on the regime, we observe markedly different
spectra. For low atom numbers, the spectrum shows separate splittings,
whereas for higher atom numbers these merge into a single
splitting. In the rest of this article, we first extend previous models of cavity
QED with multiple hyperfine levels \cite{birnbaum:06, terraciano:07,
  brennecke:09, clemens:10} to derive an effective Hamiltonian
describing our system and then detail the experiments performed and
compare them with the predictions from theory.

\section{Theory}
\label{sec:theory}

\begin{figure}[t]
  \centering
  \includegraphics[width=8.6cm]{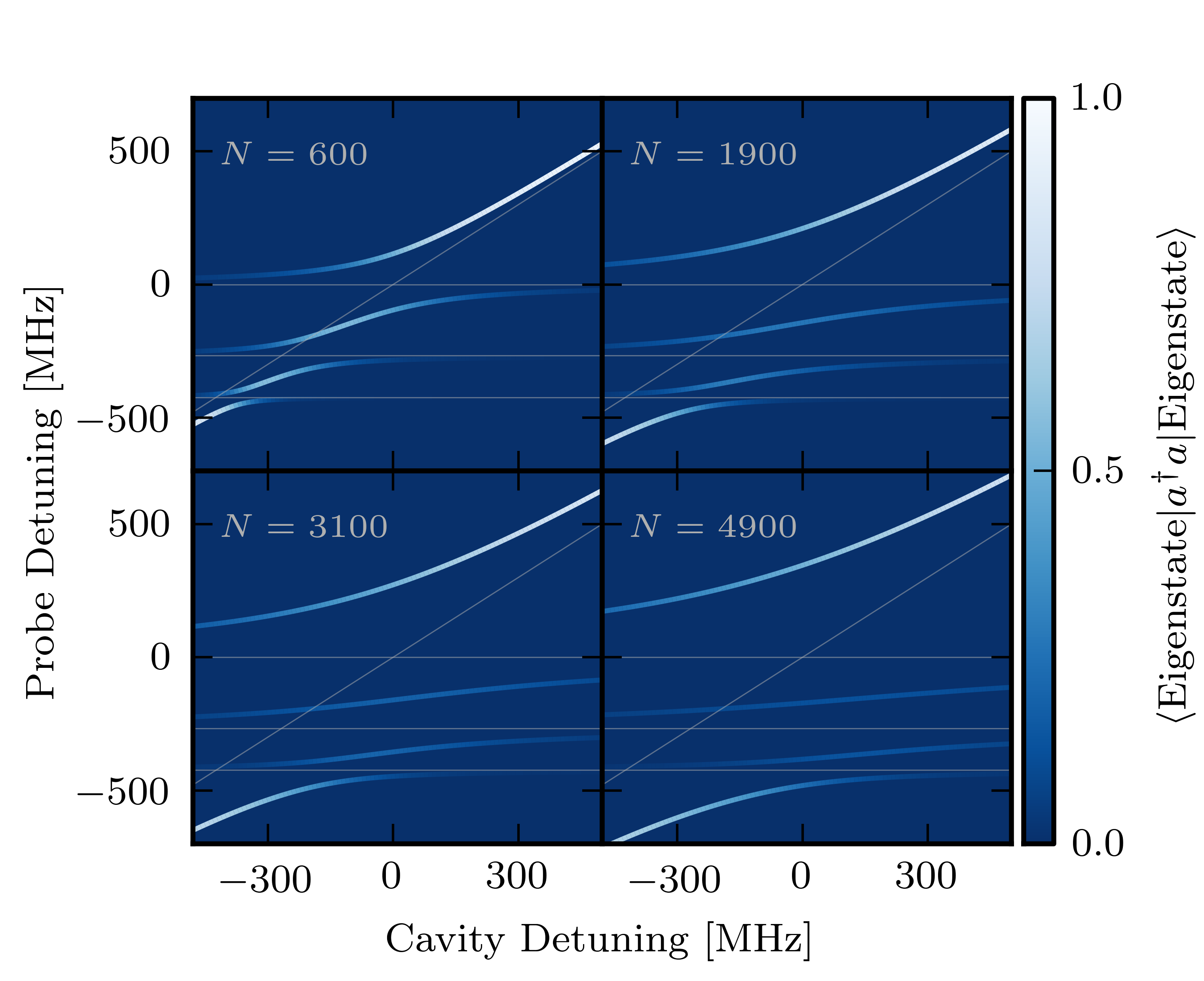}
  \caption{Theoretical transmission spectra for weak driving. Three
    avoided crossings between the energy of the bare cavity (solid diagonal
    line) and the energies of the bare atom (solid horizontal lines)
    contribute to the transmission spectrum. Transmission through the
    cavity is expected at the eigenenergies of the system. The amount
    of transmission is proportional to $\braket{a^\dagger a}$ of the
    corresponding eigenstate, as indicated by color. For low atom
    numbers the spectrum consists of three separate splittings, while
    for higher atom numbers these merge into a single
    splitting. Calculations were performed for an equal distributions
    of $m_F$ states and a $\bar{g}$ of $6.5~\MHz$. All detunings are
    with respect to the $\ket{F=2}$ to $\ket{F'=3}$ transition.}
  \label{fig:avoided-crossing}
\end{figure}

Our setup is sketched in Fig.~\ref{fig:sketch}~(a). We trap $N$ $\Rb$
atoms inside the mode of a high-finesse optical cavity. Our cavity is
a Fabry-P\'erot resonator with no discernible birefringence. Hence, it
supports two orthogonal and degenerate modes of the electromagnetic
field. By applying a magnetic field perpendicular to the optical axis
of the cavity, we ensure that these modes correspond to $\pi$ and
$\perp = (\sigma_+ + \sigma_-)/\sqrt{2} $ transitions of the atom. The
magnetic field is calculated to be 2.6 G, based on the coil
geometry. We neglect the resulting differential Zeeman shifts of
$600~\kHz$ \cite{steck:87} because they are small compared to the
cavity linewidth of 5.3 MHz.

The experiments were performed on the transitions from the $\ket{F=2}$
ground state to the $\ket{F'=1,2,3}$ excited states of the $D2$ line
\cite{steck:87}. A model Hamiltonian describing the system is

\begin{equation}
  \label{eq:full-hamiltonian}
  \begin{split}
    \HH = &\hbar \cavityfreq (a^\dagger a + b^\dagger b) + \hbar \sum_{i=1}^N
    \sum_{F'=1}^3 \Fprimefreq \ketbra{F'}{F'}_i +\\
    & \hbar \sum_{i=1}^N \sum_{F'=1}^3 \left( g_i a^\dagger \Dpi+
    g_i b^\dagger \Dperp + \text{h.c.}\right).
  \end{split}
\end{equation}
Here, $a$ and $b$ are the annihilation operators for the cavity modes
corresponding to $\pi$ and $\perp$ transitions of the atom,
$\cavityfreq$ is the resonance frequency of the cavity,
$\ketbra{F'}{F'}_i$ is the operator projecting the $i$-th atom onto
the excited state $\ket{F'}$, $\Fprimefreq$ is the frequency of the
transition from the $\ket{F=2}$ ground state to the $\ket{F'}$ excited
state, $g_i$ is the complex atom-cavity coupling constant of the
$i$-th atom and $\text{h.c.}$ denotes the Hermitian conjugate. $\Dpi$
and $\Dperp$ are the atomic dipole transition operators for the $i$-th
atom from the $\ket{F'}$ excited state to the $\ket{F=2}$ ground
state, which take into account the different coupling strengths for
the $\pi$ and $\perp$ transitions.

We follow the convention in \cite{birnbaum:06} to define the atomic
dipole transition operators for an atom interacting with different
polarizations of the light field as

\begin{equation}
  \label{eq:2}
  \begin{split}
    &D^q(F, F') =\\
    &\quad\sum_{m_F} \ket{F, m_F}\braket{F, m_F | \mu_q | F',
      m_F + q} \bra{F', m_F + q},
  \end{split}
\end{equation}
where $q=\{-1, 0, 1\}$ and $\mu_q$ is the dipole operator for
$\{\sigma_-, \pi, \sigma_+\}$-polarization, normalized such that for the
cycling transition from the $\ket{F=2, m_F=2}$ to the $\ket{F'=3,
  m_{F'} = 3}$ state $\braket{\mu} = 1$. For $\perp$-polarized light we
identify $D^\perp_i= (D^{+1}_i + D^{-1}_i) / \sqrt{2}$.

In the experiment, we probe the system with $\pi$-polarized light
after having prepared the atoms in a statistical mixture of $m_F$
states. For weak driving and large atom numbers the system behaves
like a collective spin with one ground and three excited states
coupled to the $\pi$-mode of the cavity, as indicated in
Fig.~\ref{fig:sketch}~(b). We model the system by the effective
Hamiltonian %
\begin{equation}
  \label{eq:reduced-hamiltonian}
  \begin{split}
    \HH = &\hbar \cavityfreq a^\dagger a + \hbar
    \sum_{F'=1}^3 \Fprimefreq \ketbra{F'}{F'} +\\
    &\hbar \sum_{F'=1}^3 \left( \sqrt{N} \gbar_{F'}  a^\dagger
    \ketbra{F=2}{F'} + \text{h.c.}\right),
  \end{split}
\end{equation}%
which we derive in detail in
Appendix~\ref{sec:deriv-reduc-hamilt}. The coupling of the collective
spin is enhanced by a factor $\sqrt{N}$ as compared to an effective
single atom coupling constant $\gbarFprime$ that is an average of both
the spatial dependence and the $m_F$ state dependence of the coupling
of the individual atoms. Because we are only considering the case of
weak driving, we restrict our discussion of the spectrum to the
excitation manifold with a single excitation present in the system
\cite{birnbaum:06}. In this manifold, there are three avoided
crossings between the energies of the bare cavity and the bare
atom. If the crossings are well separated, we expect three splittings
in the transmission spectrum of size $2 \sqrt{N} \gbarFprime $,
separated by the hyperfine splitting of the excited states of
$\Rb$. Solving the eigensystem for the Hamiltonian yields a more
detailed description of the spectrum. We expect to see transmission
when the probe laser frequency is resonant with the eigenenergies of
the system, while the amount of transmission is proportional to the
expectation value $\braket{a^\dagger a}$ of the corresponding
eigenstate \cite{birnbaum:06}. Numerical calculations lead us to
predict the detailed form of spectrum as summarized in
Fig.~\ref{fig:avoided-crossing}.

\section{Experiment}
\label{sec:experiment}

We verify our predictions using the experimental setup sketched in
Fig.~\ref{fig:sketch}~(a).  Its central element is a cavity made of
two coned down spherical mirrors with radii of curvature of $25~\mm$
separated by $500~\um$. The transmission per mirror is $\approx
20~\ppm$ and the losses are $\approx35~\ppm$, as estimated from
measurements of the transmission, reflection and linewidth of the
cavity. In terms of cavity QED parameters, our setup is described by
$(g, \kappa, \gamma) = 2\pi \times (9.2, 2.6 , 3.0) ~\MHz$, where $g$ is
the maximum single atom coupling constant for the $\ket{F=2, m_F=2}$ to
$\ket{F'=3, m_F=3}$ transition, $\kappa$ is the cavity field decay
rate, and $\gamma$ is the atomic dipole decay rate \cite{hunger:10,
  steck:87}.

We start the experiment by loading atoms into an optical dipole trap
in the $\ket{F=1}$ ground state, similar to our previous experiments
\cite{arnold:11}. The dipole trap is formed by a beam with a
wavelength of $1064~\nm$ and a power of $12~\W$ focused to a waist of
$25~\um$. From there the atoms are evaporatively cooled into a
transfer lattice formed by two counter-propagating beams of the same
wavelength. Each lattice beam has a maximum power of $1~\W$ and is
focused to a waist of $50~\um$ inside the cavity. By changing their
relative frequency, the atoms are transported into the cavity
\cite{kuhr:01}, which is located $9.2~\mm$ below the position of the
dipole trap. Inside the cavity, the atoms are trapped in a
two-dimensional lattice formed by the transfer lattice and an
intra-cavity beam at $808~\nm$ with a circulating power of $3~\mW$ and
a waist of $25~\um$. We also use this beam to stabilize the length of
the cavity.

Within the cavity, the atoms are repumped into the $\ket{F=2}$ ground
state and the repumping beam is left on during the rest of the
experiment. The atom-cavity system is probed through the cavity with
light linearly polarized along the axis of quantization defined by the
magnetic field. The intensity of the probe is adjusted to give a small
mean photon number $\nempty$ in the empty cavity. It is chosen to be
as close as possible to the weak probing condition $\nempty \ll 1$
while still resulting in enough signal for the experiment. The
transmission spectrum is recorded by directing the output of the
cavity onto a single photon counting module. After probing the system,
we measure the atom number via absorption imaging.

\begin{figure}[tb]
  \centering
  \includegraphics[width=8.6cm]{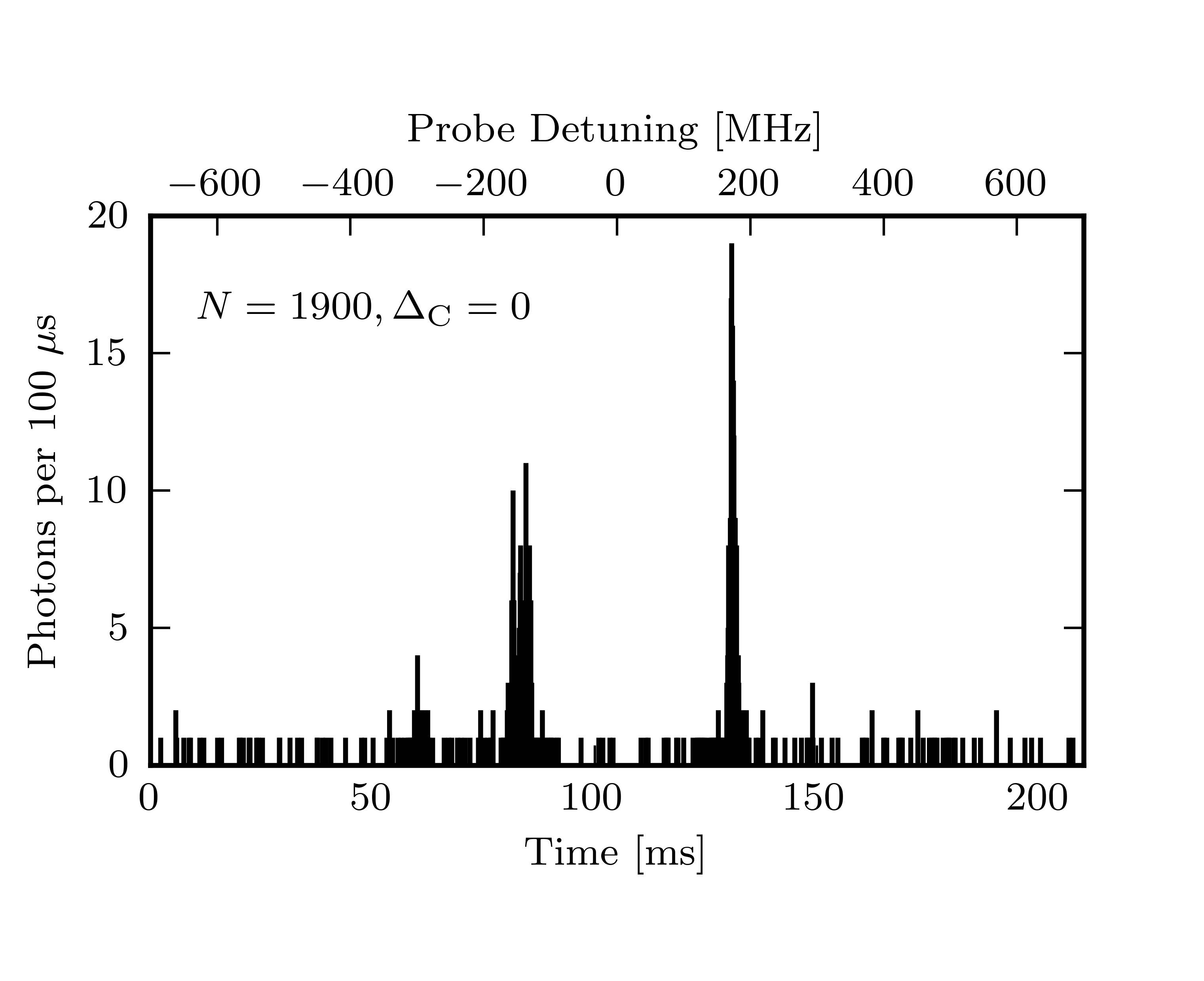}
  \caption{Cavity transmission. For a given experiment we fix the
    cavity detuning $\Delta_\text{C}$ and the atom number $N$. The
    probe laser detuning is swept from $-700~\MHz$ to $+700~\MHz$ in
    $210~\ms$, which is small compared to the measured atom number lifetime of
    3 s. At the end of the sweep the atom number is measured by
    absorption imaging. The cavity output is directed onto a single
    photon counting module and the counts are binned in $100~\us$
    intervals. All detunings are with respect to the $\ket{F=2}$ to
    $\ket{F'=3}$ transition.}
  \label{fig:single-run}
\end{figure}

For each experiment the cavity detuning $\Delta_\text{C} = \cavityfreq
- \omega_1$ is fixed and we control the number of atoms by adjusting
the power in the initial dipole trap. During the experiment the probe
laser frequency is swept over the frequency range of interest,
resulting in a transmission signal as shown in
Fig.~\ref{fig:single-run}. By repeating the experiment for different
cavity detunings and different average atom numbers, we map out the
transmission spectra, as shown in Fig.~\ref{fig:transmission}. 

\section{Conclusion}
\label{sec:conclusion}

\begin{figure}[t]
  \centering
  \includegraphics[width=8.6cm]{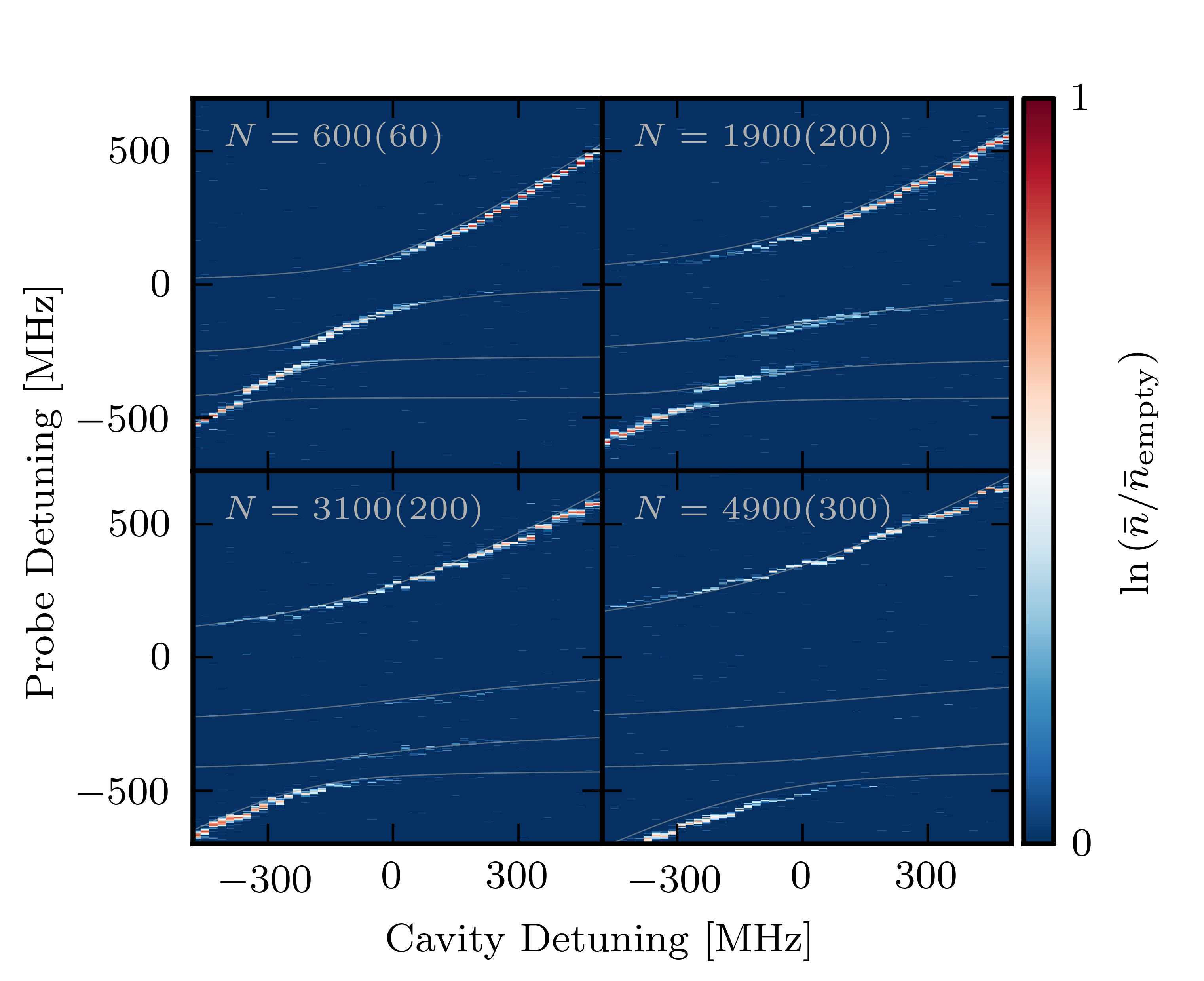}
  \caption{Experimental transmission spectra. By varying the cavity
    detuning we map out the full transmission spectra of the system
    for different average atom numbers $N$, with their standard
    deviation indicated in parentheses. The recorded transmission is
    indicated in color in terms of intra-cavity photon number
    $\bar{n}$, normalized to the average photon number inside the
    empty cavity on resonance $\nempty$, where $\nempty = 0.25$ for
    the top left and bottom right plots and $\nempty = 0.35$ for the
    top right and bottom left plots. Points with counts below 2 are
    colored as 0. The gray solid lines indicate the positions of the
    theoretical eigenenergies for the measured average atom number
    $N$, an equal distributions of $m_F$ states and a $\bar{g}$ of
    $6.5~\MHz$. All detunings are with respect to the $\ket{F=2}$ to
    $\ket{F'=3}$ transition.}
  \label{fig:transmission}
\end{figure}

To compare the measured spectra with our model predictions, we assume
equal population of both the $m_F$ states and the lattice sites of the
intra-cavity trap. For simplicity, we neglect the finite size of the
cloud along the direction transverse to the optical axis of the cavity
\footnote{For a $25~\um$ mode waist, a trap depth of $330~\uK$ and a
  typical temperature of $33~\uK$, as in our experiment, the
  transverse extend of the cloud in is $4~\um$ and reduces the
  average coupling by only 8\%}. Because the intra-cavity trap has a
different lattice spacing than the mode of the cavity, the spatial
averaging of $g_i$ gives $\bar{g} \approx \max(g_i) / \sqrt{2} =
6.5~\MHz$, as for a uniform distribution of atoms along the cavity
axis.  The agreement between the experiment and the model, as
illustrated in Fig.~\ref{fig:avoided-crossing} and
Fig.~\ref{fig:transmission}, is obtained without any free parameters.

So far, we have neglected the effect of non-neglible driving
strength. Our model assumes weak driving in order to describe the
system as a collective spin. It will fail for higher driving strength
since it predicts a strong non-linearity due to the saturation of the
single spin. However, as the atom number increases, the system
saturates at higher excitation numbers and loses the strong
nonlinearity predicted by our simple model. Another effect we have
neglected are the $m_F$ state changing processes either due to optical
pumping or due to the weak relaxation via the undriven mode of the
cavity. This is justified because the experiment was performed with a
large number of atoms in a statistical mixture of all $m_F$
states. Thus, residual $m_F$ state changing processes are not expected
to alter the distribution significantly.  In a series of related
experiments, we have started to investigate the dynamics of polarized
gases, where the atoms are optically pumped into a particular $m_F$
state prior to their interaction with the cavity. In this case, we see
evidence of $m_F$ changing processes, which in turn change the
effective coupling $\gbarFprime$ of the system. These processes lead
to complex dynamics, which will be an interesting topic for future
investigations.

In summary, we have experimentally demonstrated that the linear
response of a gas of alkali-metal atoms coupled to a high-finesse optical
cavity is well described by a collective spin with multiple levels
coupled to a single mode of the cavity. Using this system provides a
flexible testbed for collective cavity QED with multiple atomic levels
and the theoretical frame work is applicable to other physical systems
used in quantum information processing. These include hybrid systems
like alkali-metal atoms above a coplanar waveguide, collectively coupled to
linearly polarized microwave photons \cite{verdu:09}, and
nitrogen-vacancy centers coupled to a superconducting resonator, where
a similar effect in the transmission spectrum has recently been
observed \cite{kubo:10}.

\begin{acknowledgments}
  We would like to thank A.S. Parkins and F. Brennecke for
  fruitful discussions. We acknowledge the support of this work by the
  National Research Foundation and the Ministry of Education of
  Singapore, as well as by A-STAR under project No. SERC 052 123 0088.
\end{acknowledgments}

\appendix

\section{Derivation of the reduced Hamiltonian}
\label{sec:deriv-reduc-hamilt}

In order to derive the effective Hamiltonian of
Eq.~\ref{eq:reduced-hamiltonian}, we take a closer look at the
interaction part $\HHI$ of the full Hamiltonian of
Eq.~\ref{eq:full-hamiltonian}, that is,%
\begin{equation}
  \label{eq:interaction-hamiltonian}
  \HHI = \sum_{i=1}^N \sum_{F'=1}^3 \left( g_i a^\dagger \Dpi+
    g_i b^\dagger \Dperp + \text{h.c.}\right).
\end{equation}%
We will first consider the driven $\pi$-mode and then the undriven
$\perp$-mode.  Driving the $\pi$-mode of the cavity populates the
state where all the atoms are in the ground state and there is one
$\pi$-photon in the cavity,

\newcommand{\pilabel}{\pi,\mathrm{g}_{N}}
\newcommand{\zerolabel}{0,\mathrm{e}_N}
\newcommand{\perplabel}{\perp,\mathrm{g}'_N}
\newcommand{\piN}{\ensuremath{\ket{\pilabel}}}
\newcommand{\zeroN}{\ensuremath{\ket{\zerolabel}}}
\newcommand{\perpN}{\ensuremath{\ket{\perplabel}}}

\begin{equation}
  \label{eq:5}
 \piN = \ket{\pi}\prod_{i=1}^N \ket{F, m_{F}}_i.
\end{equation}%
Subsequent interaction with the cavity leads to transitions to the
Dicke state, where one of the atoms is excited and the cavity is empty,%
\begin{equation}
  \label{eq:7}
  \zeroN =\ket{0} \left(\frac{1}{\sqrt{N}} \sum_{i=1}^N \ket{F', m_{F'}}_i\prod_{j\ne
  i} \ket{F, m_{F}}_j\right).
\end{equation}%
The rate of which is collectively enhanced, i.e.,%
\begin{equation}
  \label{eq:8}
  \braket{\zerolabel|\HHI|\pilabel} = \sqrt{N} \gbarFprime.
\end{equation}%

To calculate the effective single atom coupling rate $\gbarFprime$, we
note that the coupling of the atom to the $\pi$-mode of the cavity
depends on its position relative to the cavity field. Along the cavity
axis the corresponding $g_i$ varies from maximum coupling for an atom
at an anti-node of the cavity field, to zero coupling for an atom at a
node. In the transverse direction the coupling varies as the Gaussian
field distribution of the $\mathrm{TEM}_{00}$ mode of the
Fabry-P\'erot cavity. 

In addition to position, the coupling depends on the $m_F$ state of
the atom via the Clebsch-Gordon coefficient $\ClebschPi_i$ that enters
the dipole transition operator
\begin{equation}
  \label{eq:Dpi}
  \begin{split}
    &D_i^0(F, F') =\\
    &\quad\sum_{m_F} \ket{F, m_F}_i\braket{F, m_F | \mu_0 | F',
      m_F}_i \bra{F', m_F}_i.
  \end{split}
\end{equation}%
For a product state of
$N$ atoms, each in a particular $m_F$ state, the sum in the dipole
transition operator $\Dpi_i$ has only one non-zero term for every
atom. The operator then becomes a lowering operator, taking the $i$-th
atom from the $\ket{F'}$ to the $\ket{F=2}$ state, with a prefactor
given by the Clebsch-Gordon coefficient and the spatially varying
coupling constant $g_i$, i.e.,
\begin{equation}
  \label{eq:6}
  g_i \Dpi_i = g_i \ClebschPi_i \ketbra{F=2}{F'}_i.
\end{equation}

For weak driving the ensemble behaves as a collective spin and the
effective single atom coupling constant results from both an average over
the spatial distribution of the atoms and their $m_F$ states, that is
\begin{align}
  \label{eq:4}
  \gbarFprime &= \sqrt{\frac{1}{N} \sum_{i=1}^N \abs{\braket{F,
        m_{F}|\mu_0|F', m_{F}}_i g_i}^2 } \\
  &\approx\gbar \sqrt{\sum_{m_F}p(m_F)\abs{\braket{F, m_{F}|\mu_0|F', m_{F}}}^2 }.
\end{align}
Here, $\gbar = \sqrt{\sum_i \abs{g_i}^2/N}$ is the coupling constant
resulting from spatial variations in $g_i$ and $p(m_F)$ is the
relative population in the different $m_F$ states. In the last
step, we have assumed that enough atoms are in all of the $m_F$ states
such that the atoms in each state independently average to $\gbar$.

The undriven mode of the cavity is only populated by a
transition of the excited state $\zeroN$ into the state where the
cavity holds a $\perp$-photon and the atoms are in a superposition of
one of them having changed $m_F$ states, i.e.,%
\begin{equation}
  \label{eq:9}
  \perpN = \ket{\perp} \left( \frac{1}{\sqrt{2N}} \sum_{q=\pm1}
    \ket{F, m_{F}+q}_i
      \prod_{j\ne i} \ket{F, m_{F}}_j \right).
\end{equation}
The rate at which this process occurs is not collectively enhanced
\cite{terraciano:07, clemens:10}, i.e.,%
\begin{equation}
  \label{eq:perprate}
  \braket{\perplabel|\HHI|\zerolabel} = \gbarFprime,
\end{equation}%
 even if all $m_F$ states are
macroscopically occupied \cite{brennecke:09}. Because $N\gg 1$ in the
experiment, we restrict our treatment to the interaction with the
driven $\pi$-mode of the cavity and arrive at the effective
Hamiltonian for weak driving

\begin{equation}
  \label{eq:3}
  \begin{split}
    \HH = &\hbar \cavityfreq a^\dagger a + \hbar
    \sum_{F'=1}^3 \Fprimefreq \ketbra{F'}{F'} +\\
    &\hbar \sum_{F'=1}^3 \left( \sqrt{N} \gbar_{F'}  a^\dagger
    \ketbra{F=2}{F'} + \text{h.c.}\right).
  \end{split}
\end{equation}

The treatment above is similiar to work in the area of quantum dots
studying the vacuum-Rabi splitting in the presence of inhomogenous
broadening \cite{houdre:96}.%
\bibliography{hyperfine_cqed}

\end{document}